\documentclass[fleqn,twoside]{article}
\usepackage{espcrc2}

\usepackage{graphicx}

\def\sqeeb{\ifmmode{\sqrt{s_{\protect\bf\mathrm{ee}}}}\else
  {$\sqrt{s_{\protect\bf\mathrm{ee}}}$}\fi}
\def\epem{\ifmmode{\mathrm{e}^{+}\mathrm{e}^{-}}\else
  {$\mathrm{e}^{+}\mathrm{e}^{-}$}\fi}
\def\sqee{\ifmmode{\sqrt{s_\mathrm{ee}}}\else
  {$\sqrt{s_\mathrm{ee}}$}\fi}
\def\kp{{\ifmmode{k_{\perp}}\else{$k_{\perp}$}\fi}}
\def\etmean{\ifmmode{\bar{E}^{\mathrm{jet}}_{\mathrm{T}}}\else
  {$\bar{E}^{\mathrm{jet}}_{\mathrm{T}}$}\fi}
\def\etajmean{\ifmmode{|\bar{\eta}^\mathrm{jet}|}\else 
  {$|\bar{\eta}^\mathrm{jet}|$}\fi}
\def\lxg{\ifmmode{\mathrm{log_{10}}(x_{\gamma})}\else
  {${\mathrm{log_{10}}(x_{\gamma})}$}\fi}
\def\xg{\ifmmode{x_{\gamma}}\else{${x_{\gamma}}$}\fi}
\def\xgp{\ifmmode{x_{\gamma}^+}\else{${x_{\gamma}^+}$}\fi}
\def\xgm{\ifmmode{x_{\gamma}^-}\else{${x_{\gamma}^-}$}\fi}
\def\xgpm{\ifmmode{x_{\gamma}^{\pm}}\else 
  {${x_{\gamma}^{\pm}}$}\fi}
\def\ipb{\ifmmode {\mathrm{pb}^{-1}}\else 
  {$\mathrm{pb}^{-1}$}\fi}
\def\as{\alpha_{\rm s}}
\def\ee{\ifmmode{\mbox{e}^+\mbox{e}^-}\else
  {$\mbox{e}^+\mbox{e}^-$}\fi}
\def\thetamaxp{\ifmmode{\theta_\mathrm{max}'}\else
  {$\theta_\mathrm{max}'$}\fi}
\def\pt{\ifmmode{p_\mathrm{T}}\else{$p_\mathrm{T}$}\fi}
\def\Zzero{\ifmmode{\mathrm{Z}^{0}}\else{$\mathrm{Z}^{0}$}\fi}

\def\etjet{\ifmmode{E^\mathrm{jet}_\mathrm{T}}\else 
  {$E^\mathrm{jet}_\mathrm{T}$}\fi}
\def\ptmiss{\ifmmode{{P}_{\mathrm{T,MISS}}}\else 
  {${P}_{\mathrm{T,MISS}}$}\fi}
\def\mj1h2{\ifmmode{M_{\mathrm{J1H2}}}\else 
  {$M_{\mathrm{J1H2}}$}\fi}
\def\ebeam{\ifmmode{E_{\mathrm{BEAM}}}\else 
  {$E_{\mathrm{BEAM}}$}\fi}
\def\etajet{\ifmmode{|\eta^\mathrm{jet}|}\else 
  {$|\eta^\mathrm{jet}|$}\fi}
\def\etaj{\ifmmode{\eta^\mathrm{jet}}\else 
  {$\eta^\mathrm{jet}$}\fi}
\def\etajdef{\ifmmode{\eta^\mathrm{jet} = 
    -\ln\tan(\theta^\mathrm{jet}/2)}\else{$\eta^\mathrm{jet} = 
    -\ln\tan(\theta^\mathrm{jet}/2)$}\fi}
\def\detajet{\ifmmode{|\Delta\eta^\mathrm{jet}|}\else 
  {$|\Delta\eta^\mathrm{jet}|$}\fi}
\def\costhst{\ifmmode{|\mathrm{cos}\,\Theta^{*}|}\else 
  {$|\mathrm{cos}\,\Theta^{*}|$}\fi}
\def\etajetc{\ifmmode{|\eta^\mathrm{jet}_\mathrm{cntr}|}\else 
  {$|\eta^\mathrm{jet}_\mathrm{cntr}|$}\fi}
\def\etajetf{\ifmmode{|\eta^\mathrm{jet}_\mathrm{fwd}|}\else 
  {$|\eta^\mathrm{jet}_\mathrm{fwd}|$}\fi}
\def\etah{\ifmmode {\hat{\eta}}\else{$\hat{\eta}$}\fi}

\def\dsdeta{\ifmmode{\frac{\mathrm{d}{\sigma}_{\mathrm{dijet}}}
  {\mathrm{d}\etajet}}\else
    {$\frac{\mathrm{d}{\sigma}_{\mathrm{dijet}}}
      {\mathrm{d}\etajet}$}\fi}

\def\et{\ifmmode{E_\mathrm{T}}\else{$E_\mathrm{T}$}\fi}

\def\gg{\ifmmode{\gamma\gamma}\else{$\gamma\gamma$}\fi}
\def\gsg{\ifmmode{\gamma^{\star}\gamma}\else
  {$\gamma^{\star}\gamma$}\fi}
\def\PTMIA{\ifmmode{p_\mathrm{t}^\mathrm{mi}}\else
  {$p_\mathrm{t}^\mathrm{mi}$}\fi}
\def\sas1d{SaS\,1D}

\def\grv{GRV}
\def\grvnlo{GRV\,HO}

\def\gs96nlo{GS96\,HO}
\def\lac1{LAC\,1}
\def\hadcor{\ifmmode{(1+\delta_{hadr})}\else{$(1+\delta_{hadr})$}\fi}

\newcommand{\AmS}{{\protect\the\textfont2
  A\kern-.1667em\lower.5ex\hbox{M}\kern-.125emS}}

\title{Di-Jet Production in Photon-Photon Collisions at 
$\sqeeb = 189$ to $209$~GeV}

\author{T. Wengler\address[CERN]{CERN,
EP division, 1211 Geneva 23, Switzerland} for the OPAL collaboration}
       
\begin{document}

\begin{abstract}
Di-jet production is studied in collisions of quasi-real photons at
{\epem} centre-of-mass energies {\sqee} from 189 to 209~GeV at LEP.
The data were collected with the OPAL detector. The structure of jets
is investigated and differential cross sections are measured and
compared to QCD calculations.
\vspace{1pc}
\end{abstract}

\maketitle

\section{Introduction}
\begin{figure}[htb]
\includegraphics[width=0.47\textwidth]{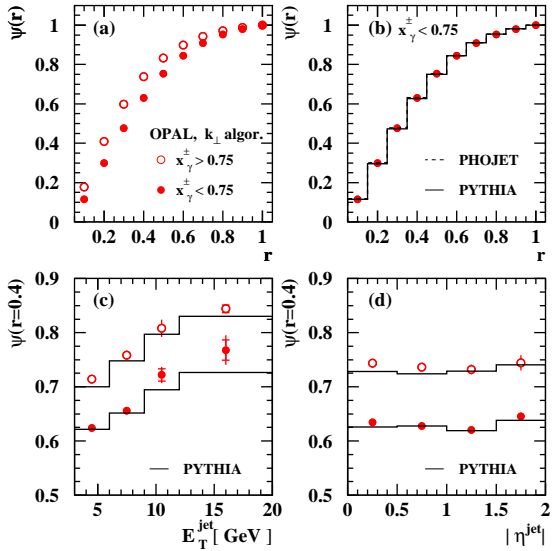}
\vspace*{-1cm}
\caption{The jet shape, $\Psi(r)$, for the two regions of 
  {\xgp}-{\xgm}-space indicated in the figure (a), and $\Psi(r)$ for
  $\xgpm < 0.75$ compared to the predictions of the LO MC generators
  PHOJET and PYTHIA (b). Figures (c) and (d) show the value of
  $\Psi(r=0.4)$ as a function of the transverse energy and
  pseudo-rapidity of the jet respectively, compared to the PYTHIA
  prediction. }
\label{fig:jsh01}
\end{figure}
\begin{figure}[htb]
\includegraphics[width=0.47\textwidth]{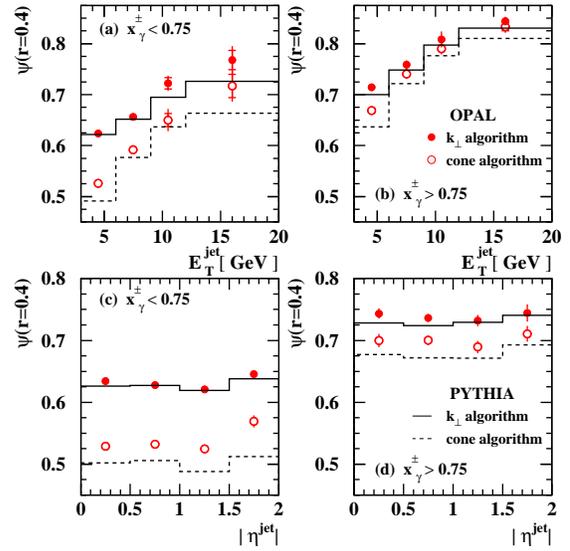}
\vspace*{-1cm}
\caption{The value of the jet shape $\Psi(r)$ at $r=0.4$ as a
  function of the jet transverse energy for ${\xgpm} < 0.75$ (a) and
  ${\xgpm} > 0.75$ (b), and as a function of the jet pseudo-rapidity
  for ${\xgpm} < 0.75$ (c) and ${\xgpm} > 0.75$ (d). In each figure
  the results obtained using the inclusive {\kp} and the cone jet
  algorithm are shown and compared to the PYTHIA prediction.}
\label{fig:jsh02}
\end{figure}

We have studied the production of di-jets in the collisions of two
quasi-real photons at an {\epem} centre-of-mass energy {\sqee} from
189 to 209~GeV, with a total integrated luminosity of
\mbox{593~\ipb} collected by the OPAL detector at LEP.  Di-jet
events are of particular interest, as the two jets can be used to
estimate the fraction of the photon momentum participating in the
hard interaction, which is a sensitive probe of the structure of
the photon. The transverse energy of the jets provides a hard scale
that allows such processes to be calculated in perturbative QCD.
Fixed order calculations at next-to-leading order (NLO) in the
strong coupling constant $\as$ for di-jet production are available
and are compared to the data, providing tests of the theory.
Leading order Monte Carlo (MC) generators are used to estimate the
importance of soft processes not included in the NLO calculation.
More details on the results presented here can be found in
\cite{bib-dijetnew}.

The {\kp}-clustering algorithm~{\cite{bib-ktclus}} is used as
opposed to the cone algorithm~{\cite{bib-coneopal}} in our previous
publications~\cite{bib-opalgg,bib-djopold} for the measurement of
the differential cross-sections, because of the advantages of this
algorithm in comparing to theoretical
calculations~{\cite{bib-ktisbest}}. The cone jet algorithm is used
to demonstrate the different structure of the cone jets compared to
jets defined by the {\kp}-clustering algorithm.

At {\ee} colliders the photons are emitted by the beam electrons
(positrons).  Most of these photons carry only a small negative
four-momentum squared, $Q^2$, and can be considered quasi-real
($Q^2 \approx 0$). The electrons are hence scattered with very
small polar angles and are not detected.  Events where one or both
scattered electrons are detected are not considered in the present
analysis.

In LO QCD, neglecting multiple parton interactions, two hard parton
jets are produced in {\gg} interactions.  In single- or
double-resolved interactions, these jets are expected
to be accompanied by one or two remnant jets.  A pair of variables,
{\xgp} and {\xgm}, can be defined~\cite{bib-LEP2} that estimate
the fraction of the photon's momentum participating in the hard
scattering:
\begin{equation}
\xgpm \equiv \frac{\displaystyle{\sum_{\rm jets=1,2}
 (E^\mathrm{jet}{\pm}p_z^\mathrm{jet})}}
 {{\displaystyle\sum_{\rm hfs}(E{\pm}p_z)}} ,
\label{eq-xgpm}
\end{equation}
where $p_z$ is the momentum component along the $z$ axis of the
detector and $E$ is the energy of the jets or objects of the
hadronic final state (hfs).  In LO, for direct events, all energy
of the event is contained in two jets, i.e.,~${\xgp}=1$ and
${\xgm}=1$, whereas for single-resolved or double-resolved events
one or both values are smaller than~1. Differential cross sections
as a function of {\xg} or in regions of {\xg} are therefore a
sensitive probe of the structure of the photon.

\section{Jet structure}
The internal structure of jets is studied using the jet shape,
which is defined as the fractional transverse jet energy contained
in a subcone of radius $r$ concentric with the jet axis, averaged
over all jets of the event sample:
\begin{equation}
\psi(r) \equiv \frac{1}{N_{\mathrm{jets}}} \sum_{\mathrm{jets}}
\frac{E\mathrm{_T^{jet}}(r)}{E\mathrm{_T^{jet}}(r=1.0)} 
\label{shape:def}
\end{equation}
with $r=\sqrt{(\Delta\eta)^2+(\Delta\phi)^2}$ and
$N_{\mathrm{jet}}$ the total number of jets analysed. Both {\kp}
and cone jets are analysed in this way. As proposed in
\cite{bib-seym}, only particles assigned to the jet by the jet
finders are considered. Events entering the jet shape distributions
are required to have at least two jets with a transverse energy
3~GeV $< {\etjet} <$ 20~GeV and a pseudo-rapidity ${\etajet} <
2$. 

In Figure~\ref{fig:jsh01}\,(a) the jet shape, $\Psi(r)$, is shown
for the {\kp} algorithm for both ${\xgpm} > 0.75$ and ${\xgpm} <
0.75$.  Here and in subsequent figures the total of statistical and
systematic uncertainties added in quadrature is shown where larger
than the marker size. The inner error bars show the statistical
errors.  The first sample is dominated by direct photon-photon
interactions and hence by quark-initiated jets. As is demonstrated
in the figure, jets in this sample are more collimated than for
small values of {\xgpm}, where the cross-section is dominated by
resolved processes and hence has a large contribution from
gluon-initiated jets. In both cases the jets become more collimated
with increasing transverse energy, as is shown in
Figure~\ref{fig:jsh01}\,(c). There is no significant dependence on
the jet pseudo-rapidity (Figure~\ref{fig:jsh01}\,(d)).  Both
PHOJET~\cite{bib-phojet} and PYTHIA~\cite{bib-pythia} give an
adequate description of the jet shapes as can be seen in
Figures~\ref{fig:jsh01}\,(b), (c), and (d). The default choices of
{\sas1d}~\cite{bib-sas} for PYTHIA and LO {\grv}~\cite{bib-grv} for
PHOJET are taken.

Figure~\ref{fig:jsh02} compares the shapes of jets defined by the cone
algorithm and the {\kp} algorithm, in each case compared to the
shape as obtained from PYTHIA. As for the {\kp}-jets, the
jets defined by the cone algorithm are more collimated in the
quark-dominated sample and always become more collimated for
increasing transverse energy, while there is no dependence on the jet
pseudo-rapidity. The cone-jets are significantly broader than the jets
defined by the {\kp} algorithm at low {\etjet}. With increasing
{\etjet}, jets become more collimated and the two jet algorithms give
similar results.  While the {\kp}-jets are well described by PYTHIA
and PHOJET, the jet shapes obtained for the cone-jets are somewhat
broader than in the data.

\section{Differential Di-jet cross-sections}
Only the {\kp} jet algorithm is used for the measurement of the
differential di-jet cross-sections. The experimental results are
compared to a perturbative QCD calculation at NLO~\cite{bib-ggnlo}
which uses the {\grvnlo} parametrisation of the parton
distribution functions of the photon~\cite{bib-grv}, and was
repeated for the kinematic conditions of the present analysis. The
renormalisation and factorisation scales are set to the maximum
$\etjet$ in the event.  The calculation was performed in the
$\overline{\mathrm{MS}}$-scheme with five light flavours and
$\Lambda^{(5)}_{\mathrm{QCD}}=130$~MeV. The average of the
hadronisation corrections estimated by PYTHIA and HERWIG have
been applied to the calculation for this comparison. 
In the figures described below the shaded band indicates the
theoretical uncertainty estimated by the quadratic sum of two
contributions: variation of the renormalisation scale by factors of
0.5 and 2 and the difference between using HERWIG or PYTHIA in
estimating the hadronisation corrections.

\begin{figure}[htb]
\includegraphics[width=0.47\textwidth]{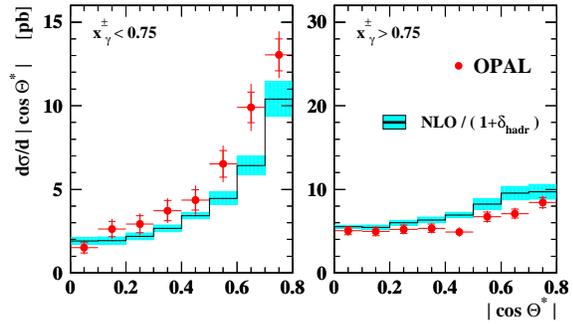}
\vspace*{-1cm}
\caption{The di-jet cross-section as a function of {\costhst} for
  the two regions in {\xgp}-{\xgm}-space indicated in the figure.}
\label{fig:costhst}
\end{figure}

\begin{figure}[htb]
\includegraphics[width=0.47\textwidth]{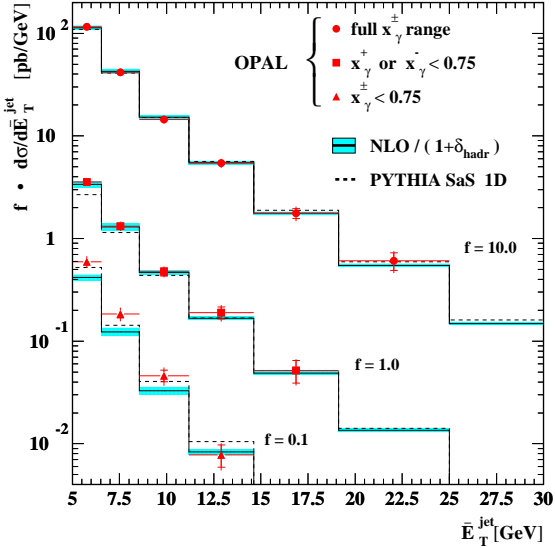}
\vspace*{-1cm}
\caption{The di-jet cross-section as a function of the mean
  transverse energy $\etmean$ of the di-jet system, for the three
  regions in {\xgp}-{\xgm}-space given in the figure.
  The factor $f$ is used to separate the three measurements in the
  figure more clearly.}
\label{fig:etmxs}
\end{figure}
\begin{figure}[htb]
\includegraphics[width=0.47\textwidth]{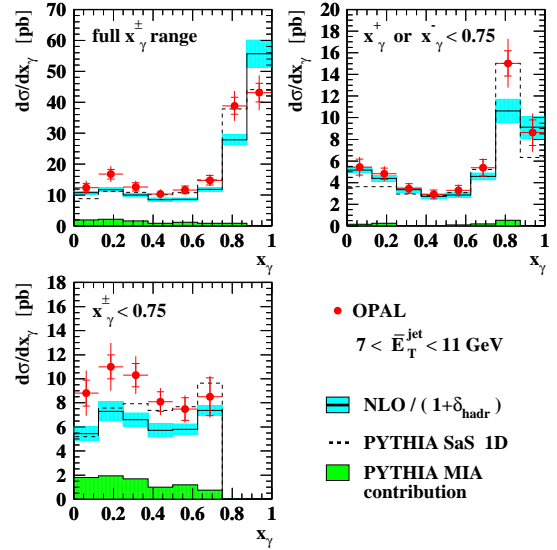}
\vspace*{-1cm}
\caption{The di-jet cross-section as a function of $\xg$ and for
the regions of the mean transverse energy $\etmean$ and {\xgpm} of
the di-jet system indicated in the figures. }
\label{fig:xgcomb}
\end{figure}

Due to the different nature of the underlying partonic process one
expects different distributions of the angle $\Theta^{*}$ between
the jet axis and the axis of the incoming partons or direct photons
in the di-jet centre-of-mass frame. The leading order direct
process $\gamma\gamma \rightarrow \mathrm{q\bar{q}}$ proceeds via
the $t$-channel exchange of a spin-$\frac{1}{2}$ quark, which
leads to an angular dependence $\propto \left( 1- \mathrm{cos}^{2}
\Theta^{*}\right)^{-1}$. In double resolved processes the sum of
all matrix elements, including a large contribution from spin-$1$
gluon exchange, leads to an approximate angular dependence
$\propto \left( 1- |\mathrm{cos}
\Theta^{*}|\right)^{-2}$~\cite{bib-cost}. The contribution of the
different processes to all resolved events depends on the
parton distribution functions of the photon.
An estimator of the angle $\Theta^{*}$ can be formed from the
pseudo-rapidities of the two jets as 
\begin{eqnarray}
\mathrm{cos}\Theta^{*} &=& \mathrm{tanh}\left(
\frac{\eta_{1}^{\mathrm{jet}} - \eta_{2}^{\mathrm{jet}}}{2}\right),
\label{eqn-cost}
\end{eqnarray}
where it is assumed that the jets are collinear in $\phi$ and have
equal transverse energy. Only {\costhst} can be measured, as the
ordering of the jets in the detector is arbitrary. To obtain an
unbiased distribution of {\costhst} the measurement needs to be
restricted to the region where the di-jet invariant mass
$M_{\mathrm{jj}} = 2\etmean/\sqrt{1-\costhst^2}$ is not influenced by
the cuts on {\etjet}~\cite{bib-djopold}. In the present
analysis a cut of $M_{\mathrm{jj}} > 15$~GeV ensures that the {\costhst}
distribution is not biased by the restrictions on {\etjet} for the
range {\costhst}$<$0.8 and ${\etajmean}=|
\left({\eta_{1}^{\mathrm{jet}} + \eta_{2}^{\mathrm{jet}}}\right)/2| <
1$ confines the measurement to the region where the detector
resolution on {\costhst} is good.

Figure~{\ref{fig:costhst}} shows the differential di-jet
cross-section as a function of {\costhst} for both ${\xgpm} > 0.75$
and ${\xgpm} < 0.75$.  The steeper rise with increasing {\costhst}
from the dominating spin-$1$ gluon exchange in the second sample is
clearly visible.  The shape of both samples is well described by
NLO QCD. For ${\xgpm} < 0.75$ the NLO calculation is about 20\,\%
below the data. It should be noted that in this region the
contribution from the underlying event, not included in the
calculation, is expected to be largest, as discussed in more detail
below. For ${\xgpm} > 0.75$ the NLO QCD prediction is about 20\,\%
above the data. While here the contribution from MIA is small, this
region is affected by rather large hadronisation corrections, which
translates into an uncertainty of the normalisation in comparing
the theoretical prediction to the data. Probably more importantly
it has been pointed out that the calculation of the cross section
becomes increasingly problematic when approaching
${\xg=1}$~\cite{bib-bertora}.

The differential di-jet cross-section as a function of the mean
transverse energy $\etmean$ of the di-jet system is shown in
Figure~{\ref{fig:etmxs}}.  At high
{\etmean} the cross-section is expected to be dominated by direct
processes, associated with the region ${\xgpm} > 0.75$.  Consequently
we observe a significantly softer spectrum for the case ${\xgpm} <
0.75$ than for the full {\xgp}-{\xgm}-space. The calculation is in
good agreement with the data for the full {\xgp}-{\xgm}-range and for
{\xgp} or {\xgm} $< 0.75$. The cross-section predicted for ${\xgpm} <
0.75$ is again below the measurement. PYTHIA 6.161 is in good
agreement with the measured distributions using the {\sas1d} parton
densities.

The three plots of Figure~{\ref{fig:xgcomb}} show
the differential cross section as a function of {\xg} for the three
regions in {\xgp}-{\xgm}-space described above. The shaded
histogram on the bottom of each of the three plots indicates the
contribution of MIA to the cross section as obtained from the
PYTHIA~{\cite{bib-pythia}} MC generator. It is evident especially
for {\xgpm}$<$1 that the MIA contribution is of about the same size
as the discrepancy between the measurement and the NLO
prediction. Furthermore it is interesting to observe that there is
next to no MIA contribution to the cross section if either {\xgp}
or {\xgm} is required to be less than one, while the sensitivity to
the photon structure at small {\xg} is retained. As one would
expect also the agreement of the NLO calculation with the
measurement is best in this case. With these measurements one is
therefore able to disentangle the hard subprocess from soft
contributions and make the firm statement that NLO perturbative QCD
is adequate to describe di-jet production in photon-photon
collisions in the regions of phase space where the calculation can
be expected to be complete and reliable, i.e. where MIA
contributions are small and for {\xg} not too close to unity. At
the same time a different sub-set of observables can be used to
study in more detail the nature of the soft processes leading to
the underlying event.


\begin{thebibliography}{9}
\bibitem{bib-dijetnew} OPAL Collaboration, G. Abbiendi et al. 
\newblock {\em Di-Jet Production in Photon-Photon collisions at
sqrt(s)ee from 189 to 209 GeV}, CERN-EP-2002-093, Submitted to Eur. Phys. J.

\bibitem{bib-ktclus}
{S.{\thinspace}Catani, Yu.L.{\thinspace}Dokshitzer, M.H.{\thinspace}Seymour and
  B.R.{\thinspace}Webber}{,}
\newblock Nucl.~Phys.~B406 (1993) 187; S.D.{\thinspace}Ellis,
  D.E.{\thinspace}Soper, Phys.~Rev.~D48 (1993) 3160.

\bibitem{bib-coneopal}
{OPAL Collaboration, R.{\thinspace}Akers et~al.}{,}
\newblock Z.~Phys.~C63 (1994) 197.

\bibitem{bib-opalgg}
{OPAL Collaboration, K.{\thinspace}Ackerstaff et~al.}{,}
\newblock Z.~Phys.~C73 (1997) 433.

\bibitem{bib-djopold}
{OPAL Collaboration, G.{\thinspace}Abbiendi et~al.}{,}
\newblock Eur.~Phys.~J.~C10 (1999) 547.

\bibitem{bib-ktisbest}
{M.{\thinspace}Wobisch and T.{\thinspace}Wengler}{,}
\newblock hep-ph/9907280; M.H.{\thinspace}Seymour, hep-ph/9707349;
  S.D.{\thinspace}Ellis, Z.{\thinspace}Kunszt and D.E.{\thinspace}Soper,
  Phys.~Rev.~Lett. 69 (1992) 3615.

\bibitem{bib-LEP2}
{L.{\thinspace}L\"onnblad and M.{\thinspace}Seymour (convenors)}{,}
\newblock {\em $\gamma\gamma$ Event Generators}, in ``Physics at LEP2'', CERN
  96-01, eds.~G.{\thinspace}Altarelli, T.{\thinspace}Sj\"ostrand and
  F.{\thinspace}Zwirner, Vol.~2 (1996) 187.

\bibitem{bib-seym}
{M.H.{\thinspace}Seymour}{,}
\newblock Nucl.~Phys.~B513 (1998) 269.

\bibitem{bib-phojet}
{R.{\thinspace}Engel}{,}
\newblock Z.~Phys.~C66 (1995) 203; R.{\thinspace}Engel and
  J.{\thinspace}Ranft, Phys.~Rev.~D54 (1996) 4244.

\bibitem{bib-pythia}
{T.{\thinspace}Sj\"ostrand}{,}
\newblock Comp.~Phys.~Comm.~82 (1994) 74; T.{\thinspace}Sj\"ostrand, LUND
  University Report, LU-TP-95-20 (1995).

\bibitem{bib-sas}
{G.A.{\thinspace}Schuler and T.{\thinspace}Sj\"ostrand}{,}
\newblock Z.~Phys.~C68 (1995) 607.

\bibitem{bib-grv}
{M.{\thinspace}Gl\"uck, E.{\thinspace}Reya and A.{\thinspace}Vogt}{,}
\newblock Phys.~Rev.~D45 (1992) 3986; M.{\thinspace}Gl\"uck,
  E.{\thinspace}Reya and A.{\thinspace}Vogt, Phys.~Rev.~D46 (1992) 1973.

\bibitem{bib-ggnlo}
{M.{\thinspace}Klasen, T.{\thinspace}Kleinwort and G.{\thinspace}Kramer}{,}
\newblock Eur.~Phys.~J.~Direct~C1 (1998) 1; B.{\thinspace}P\"otter,
  Eur.~Phys.~J.~Direct~C5 (1999) 1.

\bibitem{bib-cost}
{for example: H.{\thinspace}Kolanoski}{,}
\newblock {\em Two-Photon Physics at e$^+$e$^-$ Storage Rings}, Springer-Verlag
  (1984); B.L.{\thinspace}Combridge, J.{\thinspace}Kripfganz and
  J.{\thinspace}Ranft, Phys.~Lett.~B70 (1977) 234; D.W.{\thinspace}Duke and
  J.F.{\thinspace}Owens, Phys.~Rev.~D26 (1982) 1600.

\bibitem{bib-bertora}
L.{\thinspace}Bertora and S.{\thinspace}Frixione, these
proccedings.

\end{thebibliography}
\end{document}